\documentclass[usenatbib,fleqn]{mnras}
\usepackage{graphicx,hyperref,amsmath,amssymb,xspace}
\usepackage[dvipsnames]{xcolor}

\title[Internal dynamics of the Large Magellanic Cloud]
{Internal dynamics of the Large Magellanic Cloud from \textit{Gaia} DR2}
\author[E. Vasiliev]{Eugene Vasiliev$^{1,2}$\thanks{E-mail: eugvas@lpi.ru}\\
$^1$Institute of Astronomy, Madingley road, Cambridge, CB3 0HA, UK\\
$^2$Lebedev Physical Institute, Leninsky prospekt 53, Moscow, 119991, Russia}

\newcommand{\Gaia}{\textit{Gaia}\xspace}
\defcitealias{Helmi2018}{H18}
\defcitealias{vdMarel2014}{vdMK14}

\begin{document}
\date{Accepted 2018 September 2. Received 2018 September 2; in original form 2018 May 24}
\pagerange{L100--L104}\volume{481}\pubyear{2018}
\setcounter{page}{100}
\maketitle

\begin{abstract}
We use the proper motions (PM) of half a million red giant stars in the Large Magellanic Cloud measured by \textit{Gaia} to construct a 2d kinematic map of mean PM and its dispersion across the galaxy, out to 7~Kpc from its centre. We then explore a range of dynamical models and measure the rotation curve, mean azimuthal velocity, velocity dispersion profiles, and the orientation of the galaxy. We find that the circular velocity reaches $\sim 90$~km/s at 5~Kpc, and that the velocity dispersion ranges from $\sim30-40$~km/s in the galaxy centre to $\sim15-20$~km/s at 7~Kpc.
\end{abstract}


\section{Introduction}

\begin{figure*}
\includegraphics{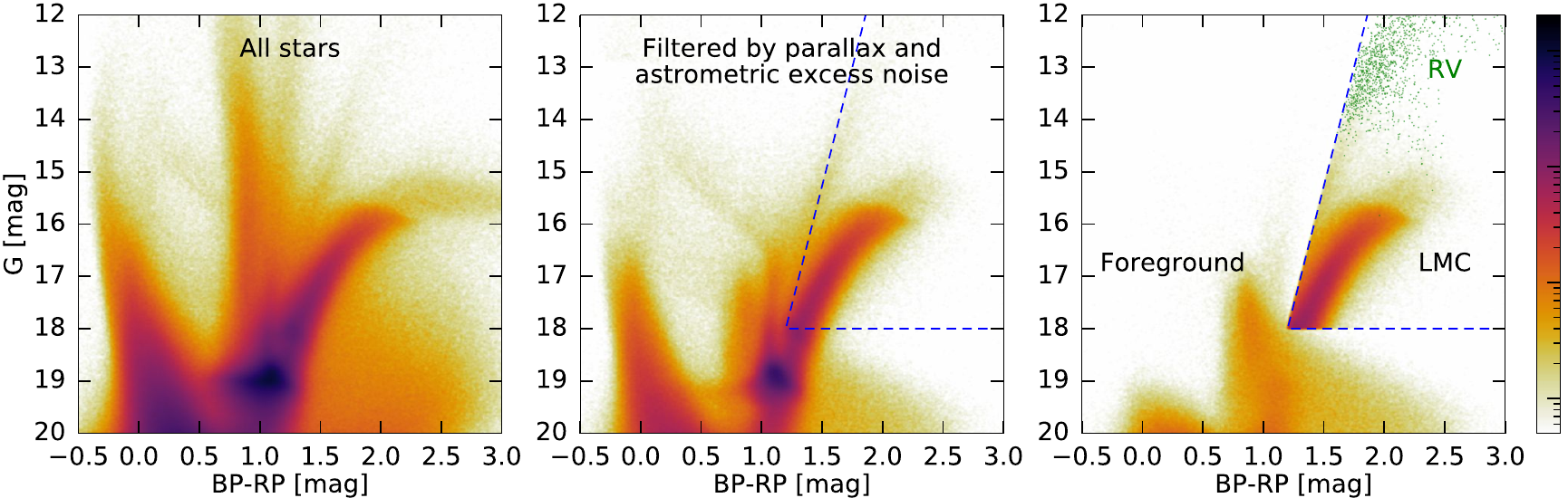}
\caption{Colour--magnitude diagram of stars within $8^\circ$ from the photometric centre of LMC.
Left panel shows all stars that have colour information in Gaia data ($\sim 10^7$). Middle panel shows the sample of stars that have parallax consistent with zero ($\varpi < 3\sigma_\varpi$) and astrometric excess noise below 0.2~mas. The region within the dashed line, occupied by red giants, is taken to be our final sample ($\sim 0.5\times10^6$ stars). Right panel illustrates that the stars which pass the parallax and astrometric excess noise cutoffs, but have proper motions inconsistent with that of LMC by more than 2~mas/yr (and hence likely belong to Milky Way), have very little overlap with our final sample; their spatial distribution is nearly uniform across the area of interest. Green dots mark stars with RV measurements.
} \label{fig:cmd}
\end{figure*}

\begin{figure}
\includegraphics{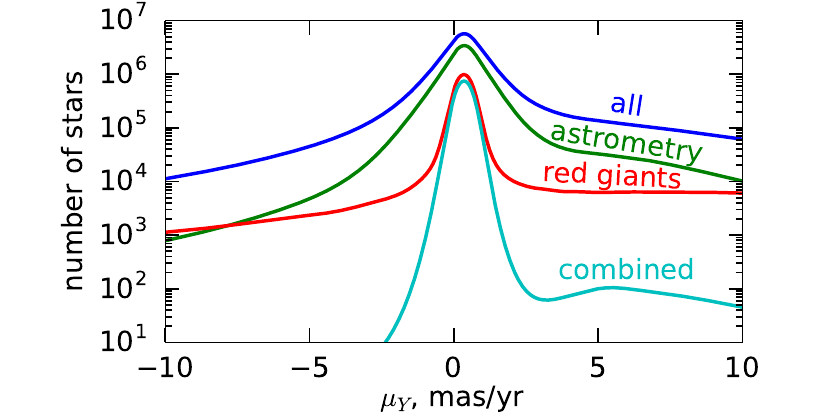}
\includegraphics{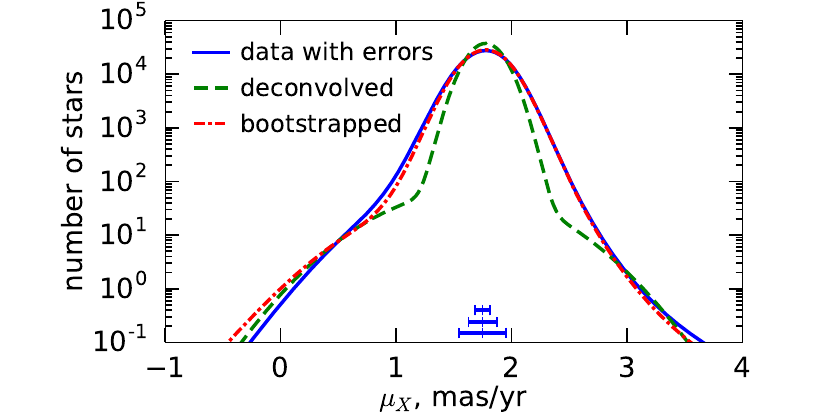} 
\caption{Histograms of proper motions $\mu_Y$ for stars with different selection criteria.\protect\\
Top panel: entire region within $8^\circ$ from the LMC centre. Blue curve shows all stars, green -- stars that pass the parallax and astrometric excess noise selection criterion, red -- stars within the region in the CMD used to pick up red giants, cyan -- the combination of both criteria. The main peak in the distribution, corresponding to the stars from LMC, has very little contamination ($<1\%$) from the broader foreground distribution.\protect\\
Bottom panel: illustration of the Extreme Deconvolution technique for inferring the intrinsic distribution of proper motions from noisy samples (simplified 1d example). Blue curve shows the actual data, green -- a two-component Gaussian mixture ($\simeq 99.5\%$ stars in the main peak) representing the intrinsic distribution, red -- the same curve convolved with errors taken from actual data points (typical errors are shown in the bottom -- 5\%, 50\% and 95\% percentiles), which follows the observed distribution. 
} \label{fig:pmhist}
\end{figure}

\begin{figure*}
\includegraphics{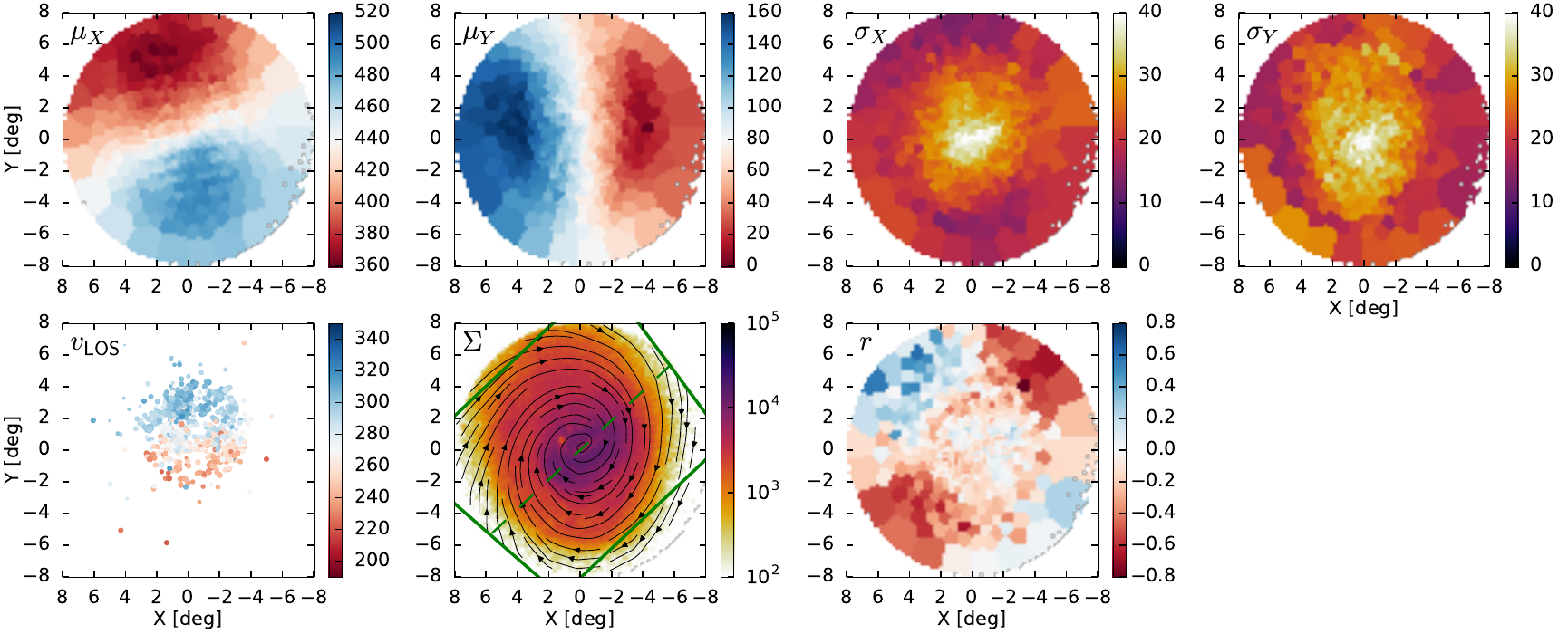}
\caption{Maps of mean PM (two top left panels), PM dispersions (two top right panels), PM correlation coefficient (bottom right), individual RV (bottom left), and density of sources per square degree (bottom centre). PM are displayed in km/s, using the multiplicative factor $1\mbox{ mas/yr}=237\mbox{ km/s}$. Bottom centre panel also depicts the streamlines of PM after subtracting their mean values, illustrating the effect of perspective shrinking due to the LMC line-of-sight motion away from the observer. The green rectangle shows the orbital plane orientation (with the top-left corner being nearer to the observer), and the dashed line shows the intersection of this plane with the sky plane (the line of nodes). The maps are available in electronic form as online supplementary material.
} \label{fig:pmobs}
\end{figure*}

The second data release (DR2) of \Gaia mission \citep{Brown2018} revolutionized the studies of kinematics of stars in the Milky Way and its satellites. The Large Magellanic Cloud (LMC), together with its smaller companion, are very prominent on the \Gaia sky map -- DR2 provides proper motion (PM) measurements for $\sim 10^7$ stars within $10^\circ$ from the LMC centre, most of which likely belong to that galaxy. 
PM measurements of individual small fields observed by the \textit{Hubble} space telescope have been available previously \citep[e.g.,][]{Kallivayalil2006,Kallivayalil2013,Piatek2008}, and were used together with radial velocity (RV) measurements of various samples of stars to construct kinematic models \citep[e.g.,][hereafter vdMK14]{vdMarel2014}. However, the \Gaia DR2 PM dataset increases the sample size by orders of magnitude and provides a uniform sky coverage. \citet[hereafter H18]{Helmi2018} used this data to construct a detailed 2d velocity map, showing a clear rotation pattern and even some deviations from it in the bar region. 

In the present paper, we use the same dataset to measure not only the mean PM, but also for the first time its dispersion. We then construct rather simple dynamical models of the LMC based on PM and $\sim1000$ stars with individually measured RV, and infer the orientation of the galaxy and the profiles of mean azimuthal velocity and velocity dispersion. These models indicate that the rotation curve reaches $\sim 90$~km/s at $5-6$~Kpc, and that the internal velocity dispersion varies in the range $15-40$~km/s; however, the residuals are still substantial, calling for more sophisticated modelling techniques.

\section{Data}

The PM dispersion of LMC stars is much more challenging to measure than its mean value, because it is sensitive to outliers and foreground contamination from the Milky Way. We therefore used a combination of several selection criteria to obtain a very clean sample:

\begin{itemize}
\item A star at the distance of LMC ($\sim 50$~kpc) would have a parallax $\varpi \sim 0.02$~mas, significantly smaller than a typical measurement error. We rejected all stars with parallaxes not consistent with zero at more than $3\sigma$ level.
\item As the astrometric model in \Gaia DR2 neglects unresolved binary stars, their PM could be substantially in error. Moreover, in dense fields such as the centre of LMC, source confusion and blending may also lead to larger errors, especially for fainter stars. This is reflected in the increased formal error bars on PM, and also in a substantially non-zero \texttt{astrometric\_excess\_noise} parameter. We therefore retained only stars with this parameter below 0.2~mas (a more stringent cutoff than in \citealt{Lindegren2018}).
\item We also removed stars with unreliable colours, mostly in crowded and highly extincted central regions, which have \texttt{phot\_bp\_rp\_excess\_factor}${}>1.3 + 0.06\,(G_{BP}-G_{RP})^2$.
\item Finally, we selected the stars from the region in the colour--magnitude diagram that corresponds to red giants (see Figure~\ref{fig:cmd}): $G$-band magnitude below 18, and $BP-RP$ colour greater than $1.2 + 0.11\,(18-G)$.
\end{itemize}
The final sample contains $\sim 0.5\times10^6$ stars, and is very clean -- the fraction of foreground contaminants is less than 1\%, as illustrated by PM histograms in Figure~\ref{fig:pmhist}, top panel. Typical PM errors in this magnitude range are $0.1-0.2$~mas/yr ($25-45$~km/s), which is smaller than the amplitude of the rotation curve ($\sim 80-90$~km/s), but comparable to the expected velocity dispersion of the galaxy.
It is therefore necessary to take into account errors in individual measurements when determining the velocity dispersion tensor. Moreover, despite the strict selection criteria, the sample still contains a small fraction of stars with measured PM that is significantly inconsistent with the bulk of the population, and would therefore bias the fitted dispersion upward.

To account for both the observational errors and contaminants, we use the Extreme Deconvolution fitting method \citep{Bovy2011}. In this approach, measurements are assumed to be drawn from an ``intrinsic'' distribution function, represented as a superposition of multivariate Gaussians, and then convolved with individual errors for each data point%
\footnote{
For a few thousand AGN sources from the AllWISE catalogue cross-matched with \Gaia, lying in the LMC region, we examined the distribution of their measured PM values normalized by the quoted uncertainties. It is best described by a normal distribution with a width 1.1, similar to the parallax uncertainties (see Figure~6 in \citealt{Lindegren2018}). We therefore increased the PM uncertainties by 10\%, and added in quadrature a systematic error of 0.04~mas/yr (typical value for spatial regions with size $\gtrsim0.2^\circ$, see Figure~15 in \citealt{Lindegren2018}), before running the Extreme Deconvolution.}.
The method aims to determine the best-fit parameters of this Gaussian mixture that maximize the likelihood of the measured values.
In our case, the data space is two-dimensional, and we use a two-component model, expecting that most of the data points come from a single (narrow) Gaussian, but a small fraction ($\lesssim 10\%$) may not be well described by this main component and instead should be attributed to a second, wider Gaussian component. The approach is illustrated in the bottom panel of Figure~\ref{fig:pmhist}: in this example, the fraction of contaminants is $<0.5\%$, but without explicitly accounting for them, we would have biased the dispersion of the main component by some $10\%$. 

In order to analyze the internal kinematics of stars, we need to consider their motion relative to the centre-of-mass of the LMC.
The latter is not a well-defined point, though, and different studies suggested positions that are more than 1$^\circ$ apart. A shift in the position of the central point produces corresponding offsets in the PM map, implying a different value for the centre-of-mass velocity, but without a proper dynamical model there is no way to determine the preferred position. As we will show later, our simple models have non-negligible systematic residuals, and therefore are not suitable to rigorously infer the position of the centre. We ran the models for several choices of the central point, obtaining largely similar results for the internal kinematics. We therefore adopted $\alpha_0 = 81^\circ, \delta_0 = -69.75^\circ$ (roughly the photometric centre of the bar) as the reference position.
After fixing the central point, we transform the coordinates and PM into a Cartesian system $X,Y$, which is an orthographic projection of the celestial sphere onto the tangent plane (defined in Equation~2 of \citetalias{Helmi2018}). Of course, since the LMC subtends a large region on the sky and a significant relative range in the line-of-sight distance, all subsequent analysis takes into account perspective effects in geometric transformations \citep[e.g.,][]{vdMarel2002}.

To construct 2d kinematic maps, we first bin the stars in the projected coordinates into square pixels with size $0.2^\circ$. The number of stars per pixel reaches $\sim 10^3$ in the central regions, but rapidly drops towards the edge of the galaxy. In order to obtain a sufficient and spatially uniform signal-to-noise, we further group pixels into $\sim 500$ Voronoi bins, using the method of \citet{Cappellari2003}. We chose a circular region with radius $8^\circ$ around the galaxy centre, because the density of stars in our sample drops rather abruptly beyond $6-7^\circ$.

Figure~\ref{fig:pmobs} (left columns), shows the map of the mean PM with a clear rotation pattern, which has been already been illustrated by \citetalias{Helmi2018} (their Figure~16). The main new result is the PM dispersion map (right columns), which has a well-defined central spike, where the values reach $\sim 0.17$~mas/yr (40~km/s), and drops to $\sim0.07-0.1$~mas/yr in the outer regions. The covariance between two components of PM also shows a clear quadrupolar pattern across the galaxy, which is unlikely to be an artefact of smaller-scale correlated measurement errors.

Uncertainties in the results are dominated by systematic errors in PM, estimated to be $\sim 0.03$~mas/yr by comparing the mean values in each pixel to a spatially-smoothed map (a similar variation is seen in Figure~17 of \citetalias{Helmi2018}). Uncertainties from the Extreme Deconvolution method (estimated by bootstrapping, i.e., choosing different subsets of data points) and from the Voronoi binning are both at the level $0.015$~mas/yr. Changing the selection criteria on the input sample (in particular, the magnitude range) has little influence on the PM dispersion: for instance, taking the upper limit at $G=17$ halves the sample size and changes the dispersion by $\lesssim 0.01$~mas/yr.
For reference, $0.01$~mas/yr is 2.4~km/s at the distance of LMC.

\section{Model}

\begin{figure*}
\includegraphics{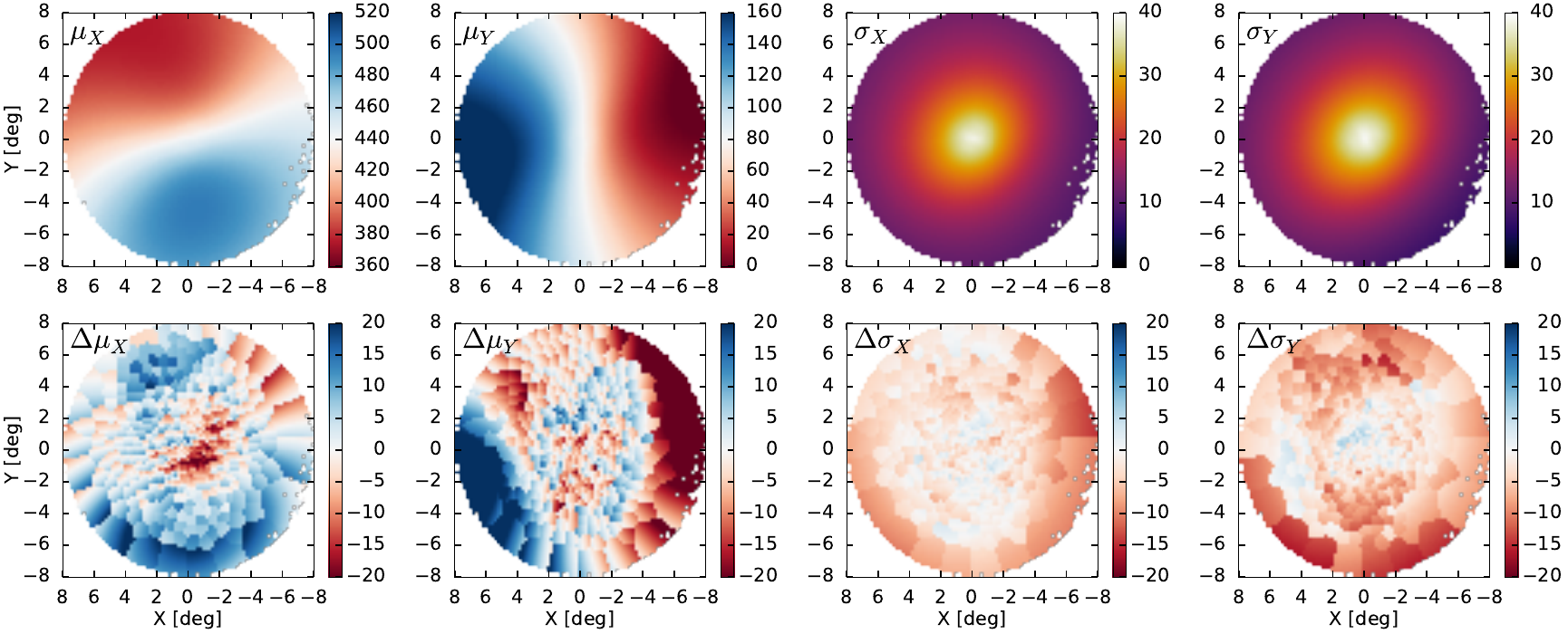}
\caption{Maps of mean PM and their dispersions for the best-fit JAM models (top row) and the residuals (bottom row). The units are again km/s as in the previous figure.
} \label{fig:pmmodel}
\end{figure*}

\begin{figure}
\includegraphics{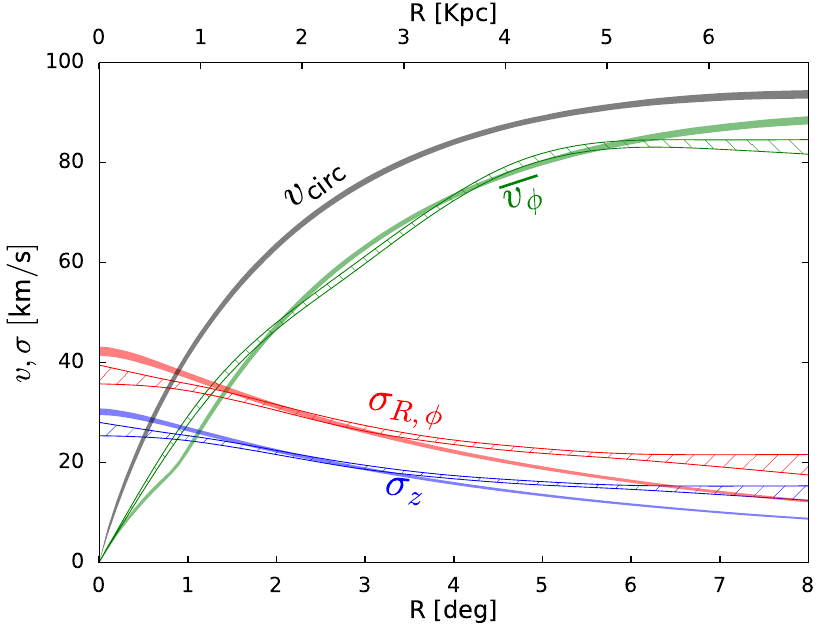}
\caption{Internal dynamics of LMC recovered by JAM (shaded) and thin-disc (hatched) models.
Red and blue are the radial and vertical velocity dispersions; green is the mean azimuthal velocity; gray is the rotation curve (circular velocity in the $z=0$ plane) in JAM.
\vspace*{-4mm}
} \label{fig:internal_dynamics}
\end{figure}

To obtain qualitative understanding of the internal dynamics of LMC, we use two different, rather simple, modelling approaches. We only take the Voronoi-binned kinematic maps of mean PM and their dispersions, plus $\sim1000$ individual radial velocity measurements, as our input data, without using any information on the density profile (the latter is known to approximately follow an exponential profile with scale radius in the range $1.3-1.7$~kpc, e.g., \citealt{vdMarel2001,Balbinot2015,Choi2018}).

The first approach is purely empirical, not based on any dynamical arguments. We assume that the stars are moving in a geometrically thin axisymmetric disc, and at each radius their velocity distribution in cylindrical coordinates $(R,z,\phi)$ is represented by a Gaussian with anisotropic dispersion tensor ($\sigma_R=\sigma_\phi=(1-\beta)^{-1/2}\sigma_z$) and mean streaming velocity $\overline{v_\phi}$. 
We adopt a cubic spline representation for the radial profiles of these functions, parametrized by their values at several control points, and assume a constant anisotropy coefficient $\beta$.
Of course, in reality a non-zero dispersion implies finite vertical thickness, and the radial variation of other quantities also cannot be arbitrary; we ignore these inconsistencies, as the primary goal of this approach is to explore the trends in the data. It is similar to a thin-disc model used in many previous studies, but is more flexible (does not make the assumption of a solid-body rotation), and additionally allows to fit for the PM dispersion.

The other method is based on the axisymmetric Jeans equations, in the form known as the Jeans Anisotropic Model (JAM) \citep{Cappellari2008}. We assume that the velocity ellipsoid at each point in the meridional ($R,z$) plane is aligned with cylindrical coordinates, and that the ratio  $\sigma_z/\sigma_R = \sqrt{1-\beta}$ is constant; moreover, we distribute the kinetic energy of azimuthal motion between ordered and random components in such a way that $\sigma_\phi=\sigma_R$. Under these assumptions, the two-dimensional ($R,z$) profiles of all kinematic quantities are uniquely determined by the total gravitational potential and the density profile of tracer stars. We parametrize the latter by a radially-exponential, vertically-isothermal disc profile, and assume that the total potential consists of the stellar disc plus a spherical halo having a double power-law profile \citep{Zhao1996}, with outer slope fixed to 3 and an arbitrary inner slope $\gamma$. In total, this model has seven free parameters (disc and halo masses, scale radii, disc scale height, halo inner slope, and the velocity anisotropy coefficient $\beta$ of the Jeans model). The observed mean PM and RV and their dispersions are obtained by integrating the intrinsic values along the line of sight. Our implementation is more general than those of \citet{Cappellari2008} and \citet{Watkins2013}, in that it allows to use arbitrary density and potential profiles (not only Multi-Gaussian expansions), and performs appropriate perspective corrections for sources at finite distances. 
As in the latter study, we evaluate the likelihood of the model against individual RV measurements, but for the PM and its dispersion we use the Voronoi-binned values obtained by the Extreme Deconvolution technique. 

In both approaches we have five additional parameters, describing the orientation of LMC disc plane (the inclination angle $i$ and the position angle of the line of nodes $\Omega$, which measures the direction to intersection between disc and image planes from the vertical axis), and velocity vector of its centre of mass in the heliocentric reference frame $v_\mathrm{CM}$. We fix the distance to 50~kpc \citep{Freedman2001}.

To explore the parameter space, we use the Markov Chain Monte Carlo (MCMC) code \textsc{Emcee} \citep{EMCEE}, running it for several thousand steps while visually checking the convergence.
We performed several validation tests by producing mock datasets (two-dimensional maps of mean PM and its dispersion, without added noise) from $N$-body models (D.Erkal, priv.comm.), and both approaches were able to recover the orientation, centre-of-mass velocity, and internal kinematics of these models rather well. 
Even though the MCMC approach provides both the best-fit parameters and their error estimates, we do not consider the latter to reflect the true uncertainties. First, because of the hard-to-control systematic errors in the input data, we used constant error bars (0.03~mas/yr) for all Voronoi bins; second, the residuals clearly indicate that the models are far from being a good match to the data, hence the formal error bars would have little sense. Instead we use the difference between the two modelling approaches, and varied the selection criteria in the input dataset, to gauge the systematic errors.

The best-fit kinematic maps are presented in Figure~\ref{fig:pmmodel} for the JAM method; the thin-disc model produces broadly similar results. Neither method provides particularly good fits to the actual data: there are systematic differences in the rotation field at the level $\sim 10$~km/s (especially in the bar region, similarly to Figure~17 in \citetalias{Helmi2018}), and in the outer region the PM dispersion is too low, while the amplitude of rotation is too high. This is not surprising, given the simplified nature of our models, while in reality the LMC is neither axisymmetric nor in perfect equilibrium, and possibly warped.
Nevertheless, the similarity of the derived one-dimensional profiles of mean velocity and its dispersion, shown in Figure~\ref{fig:internal_dynamics}, between the two methods is encouraging. 

The mean azimuthal velocity profile $\overline{v_\phi}(R)$ is very similar to Figure~18 in \citetalias{Helmi2018} and agrees well with \citet{Olsen2011}, \citetalias{vdMarel2014}, rising to $\sim 80$~km/s at the distance of 5~Kpc. The radial and azimuthal velocity dispersions $\sigma_{R,\phi}$ range from 20 to 40 km/s. This places the LMC into the category of fast rotators, according to the classification scheme of \citet{Cappellari2007}. 
The circular velocity implied by the JAM potential model is higher than $\overline{v_\phi}$ by $\sim10$~km/s and reaches 90~km/s at 5~Kpc, corresponding to an enclosed mass of $\sim10^{10}\,M_\odot$ (the relative contribution of disc and halo is not well constrained, and hence we cannot reliably extrapolate to large radii to infer the total mass).

The inclination of the LMC disc $i \sim 32-35^\circ$ and the kinematic position angle $\Omega \sim 130-135^\circ$ are in the range inferred in other studies (e.g., Table~1 in \citealt{Subramanian2013}). 
Since the galaxy orientation is closer to face-on than edge-on, the line-of-sight velocity dispersion is mainly dominated by $\sigma_z$, whereas the PM dispersion is mostly determined by $\sigma_{R,\phi}$. 
Hence a radially-anisotropic velocity distribution (in both approaches $\beta\simeq 0.5$, at the upper end of our allowed range) naturally leads to a higher PM dispersion than the RV dispersion. The latter lies in the range $\sim 15-30$~km/s, similar to other studies (e.g., \citealt{vdMarel2002}, \citealt{Cole2005}, \citetalias{vdMarel2014}, \citealt{Song2017}).
We caution that the sample of stars with \Gaia RV measurements contains relatively young red supergiants in several star-forming regions (bottom left panel in Figure~\ref{fig:pmobs}), whose kinematics likely differs from the bulk of the population. Hence we refrain from making conclusive statements about the disc thickness and vertical profile.
The best-fit line-of-sight velocity of the LMC centre-of-mass is $\sim 265-270$~km/s, comparable to the range $255-265$~km/s derived from other RV surveys, and the variation of mean RV across the image plane is qualitatively consistent with Figure~3 in \citetalias{vdMarel2014}.

\section{Summary}

In this paper, we used the ground-breaking \Gaia DR2 dataset to construct 2d maps of mean PM and its dispersion within $8^\circ$ from the LMC centre. Using a combination of selection criteria, we obtained a very clean sample of LMC stars, and inferred the intrinsic PM distribution from binned data with the Extreme Deconvolution technique, fully taking into account the observational errors and contamination. We then constructed rather simple dynamical models of LMC using two different approaches, and measured the intrinsic profiles of mean velocity, its dispersion, and circular velocity as functions of radius. Despite their simplicity, these models provide useful insights into the LMC structure, suggesting a substantial in-plane velocity dispersion in the central region and a radially anisotropic dispersion tensor. This work could be improved in many ways, in particular, taking into account more observational information on the density profile and RV, and relaxing the assumption of axisymmetry and cylindrical alignment of velocity ellipsoid on the modelling side.

I thank V.~Belokurov, D.~Erkal, N.~W.~Evans, N.~Hambly, S.~Koposov, and the anonymous referee for valuable suggestions.
This work relied exclusively on the data from the European Space Agency (ESA) mission \Gaia (\url{https://www.cosmos.esa.int/gaia}), processed by the \Gaia Data Processing and Analysis Consortium. 
This work was supported by the European Research council under the 7th Framework programme  (grant No.\ 308024). 

\end{document}